\documentclass[aps,amsmath, prb,twocolumn,superscriptaddress]{revtex4-2}
\usepackage{lineno}
\usepackage{graphicx}
\usepackage{dcolumn}
\usepackage{bm}
\usepackage{lipsum}
\usepackage{color}
\usepackage{amsfonts,eucal,mathrsfs,amssymb}
\DeclareMathAlphabet{\mathscrbf}{OMS}{mdugm}{b}{n}

\usepackage[table]{xcolor}
\graphicspath{{./figs/}}
\usepackage{multirow}
\usepackage[breaklinks=true,unicode=true,urlcolor = blue,colorlinks = true,citecolor = blue,linkcolor = blue]{hyperref}
\definecolor{commentpurple}{RGB}{191, 85, 236}
\renewcommand{\vec}[1]{\bm{#1}}
\newcommand{\dd}{\mathrm{d}}

\begin{document}

\preprint{}
%
%
\title{Thermal spin transport in easy-planar $d$-wave altermagnets controlled by magnetic field}
%
\author{Yuliia I. Gusieva}
\affiliation{G.V. Kurdyumov Institute for Metal Physics of the NAS of Ukraine, 03142 Kyiv, Ukraine}
\affiliation{Kyiv Academic University, 03142 Kyiv, Ukraine}

\author{Kostiantyn V. Yershov}
\affiliation{Institute for Theoretical Solid State Physics, Leibniz Institute for Solid State and Materials Research Dresden, D-01069 Dresden, Germany}
\affiliation{Bogolyubov Institute for Theoretical Physics of the National Academy of Sciences of Ukraine, 03143 Kyiv, Ukraine}
\author{Jeroen van den Brink}
\affiliation{Institute for Theoretical Solid State Physics, Leibniz Institute for Solid State and Materials Research Dresden, D-01069 Dresden, Germany}
\affiliation{Institute of Theoretical Physics and W{\"u}rzburg-Dresden  Cluster of Excellence {\it ct.qmat}, Technische Universit{\"a}t Dresden, 01062 Dresden, Germany} 
\author{Volodymyr P. Kravchuk}
\email{v.kravchuk@ifw-dresden.de}
\affiliation{Institute for Theoretical Solid State Physics, Leibniz Institute for Solid State and Materials Research Dresden, D-01069 Dresden, Germany}
\affiliation{Bogolyubov Institute for Theoretical Physics of the National Academy of Sciences of Ukraine, 03143 Kyiv, Ukraine}

\begin{abstract}

Altermagnets constitute a novel class of collinear spin-compensated materials in which magnon branches are spin-split even in the non-relativistic limit. The latter is the result of a more complex symmetry operation (compared to conventional antiferromagnets) that connects the two sublattices. Altermagnetic splitting strongly affects the magnon transport properties, leading, in particular, to the thermal magnon splitter effect in \textit{easy-axial} $d$-wave altermagnets. Whether this effect also appears in \emph{easy-planar} systems is not obvious, since the magnon branches do not carry a constant magnetic moment in this case. Here, we consider the \textit{easy-planar} $d$-wave altermagnet of a rutile-type, e.g., NiF$_2$, and demonstrate the emergence of the altermagnetically generated magnon magnetic moment, which is momentum-dependent and possesses $d$-wave symmetry. This then leads to the spin-splitter effect, i.e., the emergence of a magnon-driven spin current (the flow of magnetic moment) in response to the applied temperature gradient. We also demonstrate that the corresponding thermal spin conductivity can be effectively tuned by an external magnetic field perpendicular to the easy plane.


\end{abstract}

\maketitle

\section{Introduction}

Modern spintronics is undergoing a period of rapid evolution, offering alternative paradigms for information processing by utilizing the spin degree of freedom. Traditionally, electrons have been the primary carriers of spin currents; however, this restricts the choice of materials to conductors and leads to significant energy dissipation due to Joule heating \cite{Zutic04,Wolf01,Baltz18,Jungwirth16}. A vast number of promising magnetic materials are insulators, many of which belong to the class of antiferromagnets (AFMs) \cite{Zutic04,Baltz18,Jungwirth16}. In conventional collinear AFMs, magnon branches are degenerate and carry either opposite spins or are not spin-polarized. Consequently, a net magnon spin current does not arise naturally; applying the temperature gradient to such systems, one merely excites thermal transport without leading to a directional transport of magnetic moment in the absence of an external field \cite{Rezende19,Gomonay14}.

The situation changes fundamentally with the discovery of altermagnetism (A$\ell$M) — the third fundamental class of collinear magnets \cite{Smejkal22a,Smejkal22b}. Due to a specific crystal symmetry where sublattices with opposite spins are connected also by a rotation or mirror reflection, in addition to translation and time reversal, altermagnets~(A$\ell$Ms) exhibit non-relativistic band splitting, which is characteristic of both electron and magnon spectra \cite{Smejkal20,Smejkal22,Hayami20,Yuan20}. In easy-axial materials, this splitting ensures that each magnon branch carries its own fixed magnetic moment, enabling the generation of energy-efficient, thermal magnon spin currents -- the thermal magnon spin-splitter effect~\cite{Yershov24b,Weissenhofer24,Cui23a,Yang26b}.

The magnon spin-transport phenomena have already become the subject of intense investigation. So far the research on $d$-wave A$\ell$Ms has primarily focused on rutile-structured compounds such as RuO$_2$ \cite{Smejkal23,Zhou24,Li25b,Adamantopoulos24}, MnF$_2$~\cite{Bhowal24,Faure25,Yuan20}, FeF$_2$~\cite{Sears26}, and CoF$_2$~\cite{Adamantopoulos24,Yu24} with magnetic easy axis. In these systems, the emergence of magnon Nernst and Seebeck spin currents has been demonstrated, alongside predictions of the magnon orbital angular momentum \cite{Cui23a,Wu_2025, Weibenhofer26,Yang26,Cui23a,Weissenhofer24}. 
For the extensively studied easy-axial configurations of these rutiles, the $d$-wave splitting of magnon branches induces thermally driven magnon spin currents (flow of the magnetic moment), where the magnon magnetic moment $\vec{\mu}_{\nu}(\vec{k})=-\partial_{\vec{B}}\mathcal{E}_{\vec{k},\nu}$ remains a fixed quantity parallel to the Néel vector \cite{Yershov24b}. Here $\mathcal{E}_{\vec{k},\nu}$ is the energy of the magnon of $\nu$-th branch with the wave vector $\vec{k}$, and $\vec{B}$ is the applied magnetic field. The relation $\mu_{\nu}(\vec{k})$ can be obtained by considering the total magnetic moment $\vec{M}$ of the Bose-gas of magnons as a thermodynamic variable $\vec{M}=-(\partial F/\partial\vec{B})_{T,V}$ with $F$ being the Helmholtz free energy~\cite{Ashcroft76,Aharoni96,Neumann20,Yershov24b}.

In $g$-wave structures, such as CrSb and MnTe, it has been established that the properties of the magnon magnetic moment crucially depend on the magnetocrystalline anisotropy~\cite{Kravchuk25a}. While in easy-axial CrSb the magnon moment is not influenced by A$\ell$M and is a momentum-independent constant within a given branch, in easy-planar MnTe, it is induced by the A$\ell$M and is momentum-dependent~\cite{Kravchuk25a}. The emergence of the altermagnetically induced magnon magnetic moment in MnTe, however, does not lead to thermal spin transport. In the linear regime (with respect to the temperature gradient), the thermal spin conductivity vanishes due to the high symmetry of $g$-wave A$\ell$Ms. 

Here we combine the robust efficiency of the magnon spin-current generation in the lower-symmetry $d$-wave rutiles~\cite{Yershov24b,Wu_2025,Yang26} with the dynamical tunability afforded by an altermagnetically-induced $\mu_\nu(\vec{k})$ in the \textit{easy-planar} structures. As a case study, we consider the easy-planar rutile NiF$_2$. We investigate the thermal spin-splitter effect in this compound and demonstrate that the thermal spin conductivity can be effectively controlled by a magnetic field applied perpendicular to the easy plane. 
We show that, similarly to $g$-wave MnTe, the easy-planar ground state NiF$_2$, combined with the $d$-wave A$\ell$M, induces a sublattice-precession inequivalence leading to the emergence of a $k$-dependent magnon magnetic moment. However, in contrast to MnTe, in NiF$_2$ such polarized magnons give rise to a linear thermal spin-splitter effect. 


\section{Model and its ground state}\label{sec:model}
\begin{figure}
    \centering
    \includegraphics[width=\columnwidth]{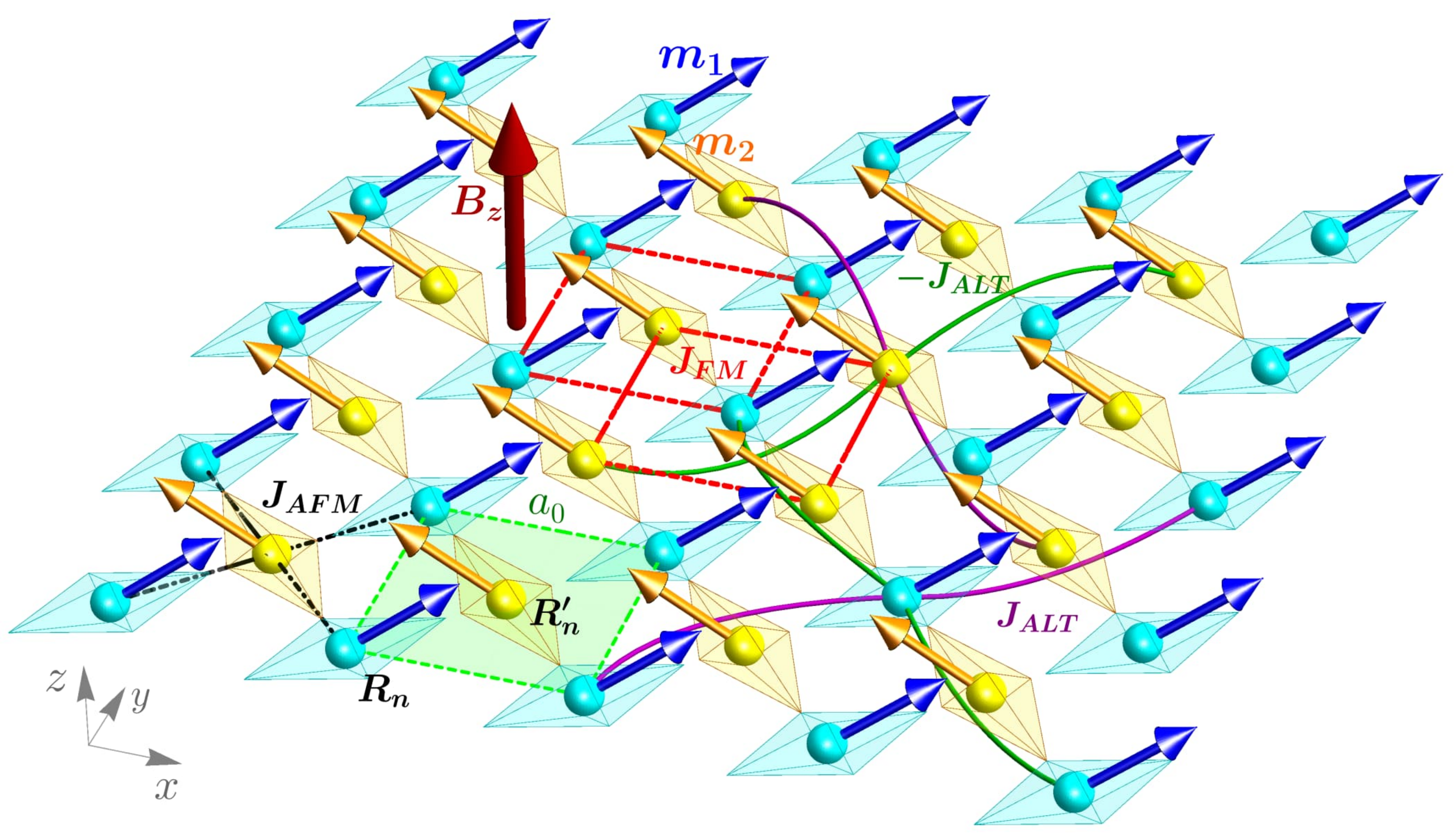}
    \caption{ Schematic crystal structure and exchange interactions in a 2D $d$-wave altermagnet of rutile type. The model consists of two magnetic sublattices, $A$ (cyan spheres, $\vec{R}_{\vec{n}}$) and $B$ (yellow spheres, $\vec{R}_{\vec{n}}'$), with antiparallel (for $B_z=0$) moments $\vec{m}_1$ and $\vec{m}_2$. Several Heisenberg exchange interactions are considered: the antiferromagnetic exchange $J_{\textsc{afm}}$ (black dotted lines) couples nearest-neighbors of different sublattices; the ferromagnetic exchange $J_{\textsc{fm}}$ (red dashed lines) couples nearest-neighbors within the same sublattice. The altermagnetic properties are modeled by additional superexchange bonds $J_{\textsc{alt}}$ and $-J_{\textsc{alt}}$ shown by the curved magenta and green lines, respectively. Note that the altermagnetic bonds are rotated by 90$^\circ$ in different sublattices. The elongated cyan and yellow octahedra represent the local ligand environments of sublattices $A$ and $B$, respectively. The magnetic unit cell is a square with the side length of $a_0$, it is shown in green.}\label{fig:model}
\end{figure}

The magnetic structure of the rutile-type altermagnet consists of ferromagnetic (001) layers coupled antiferromagnetically. We employ a 2D approximation consisting of two layers, see Fig.~\ref{fig:model}. The corresponding two sub-lattices are represented by the yellow and cyan spheres with unit magnetic moments $\vec{m}_1(\vec{R}_{\vec{n}})$ and $\vec{m}_2(\vec{R}'_{\vec{n}})$, respectively. Here $\vec{R}_{\vec{n}}=a_0(n_x\vec{e}_x+n_y\vec{e}_y)$, with $n_x,\;n_y\in\mathbb{Z}$, and $\vec{R}'_{\vec{n}}=\vec{R}_{\vec{n}}+\frac{a_0}{2}(\vec{e}_x+\vec{e}_y)$ determines positions of of the magnetic moments in two sublattices. Here, $a_0$ is the size of the magnetic unit cell, which is also the distance between the nearest neighbors within one sub-lattice. The system is described by the Hamiltonian $\mathcal{H} = \mathcal{H}_{\textsc{afm}} + \mathcal{H}_{\textsc{fm}} + \mathcal{H}_{\textsc{alt}} + \mathcal{H}_{\textsc{an}} + \mathcal{H}_{\textsc{z}}$. Here, $\mathcal{H}_{\textsc{afm}}$ and $\mathcal{H}_{\textsc{fm}}$ represent the inter-sublattice antiferromagnetic ($J_{\textsc{afm}}$, black dotted lines) and intra-sublattice ferromagnetic ($J_{\textsc{fm}}$, red dotted lines) Heisenberg exchanges, respectively. The explicit form of $\mathcal{H}_{\textsc{afm}}$ and $\mathcal{H}_{\textsc{fm}}$ is presented in Appendix~\ref{app:model}.


The key feature of this model is the inclusion of non-magnetic surrounding of the magnetic atoms, which is schematically shown by octahedra (light shapes) in Fig.~\ref{fig:model}
In rutile structures, the non-magnetic surrounding is rotated by 90$^\circ$ relative to each other for the two sublattices. This rotation breaks the combined time-reversal and translation symmetries, leading to a spin splitting of the bands with $d$-wave symmetry in momentum space. In the magnetic Hamiltonian, the symmetry of the spatial distribution of the non-magnetic surroundings is taken into account by means of the additional superexchange interactions, whose bonds are shown by the magenta and green lines in Fig.~\ref{fig:model}. Note that the bonds of different strengths are rotated by 90$^\circ$ in different sublattices. Since the altermagnetic effects are determined by the difference of the perpendicular bonds (magenta and green), we denote their strengths as $J_{\textsc{alt}}$ and $-J_{\textsc{alt}}$. In other words, we use a simplified model of 2D rutile here, previously proposed in Ref.~\onlinecite{Yershov24b}. The described interactions constitute the contribution $\mathcal{H}_{\textsc{alt}}$, whose explicit form is presented in Appendix~\ref{app:model}.


Additionally, we include the easy-plane anisotropy $\mathcal{H}_{\textsc{an}} = K\left[\sum_{\vec{R}_{\vec{n}}}m_{1z}^2(\vec{R}_{\vec{n}})+\sum_{\vec{R'}_{\vec{n}}}m_{2z}^2(\vec{R'}_{\vec{n}})\right]$ with $K > 0$, which tends to order the magnetic moments in $xy$-plane.  
The interaction with an external field $\vec{B}=B\vec{e}_z$   (large red arrow labeled $B_z$ in Fig.~\ref{fig:model}), is captured by the Zeeman term $\mathcal{H}_{\textsc{z}} = -B_z\mu_s\left[\sum_{\vec{R}_{\vec{n}}}m_{1z}(\vec{R}_{\vec{n}})+\sum_{\vec{R}'_{\vec{n}}}m_{2z}(\vec{R}'_{\vec{n}})\right]$, where $\mu_s$ is magnitude of each magnetic moment. The applied field leads to the canting of the magnetic moments, resulting in net magnetization along the field. 

In terms of the spherical angles, the ground state of each sublattice is $\vec{m}^0_\nu=\sin\theta_\nu(\vec{e}_x\cos\phi_\nu+\vec{e}_y\sin\phi_\nu)+\cos\theta_\nu\vec{e}_z$ with $\nu=1,2$. Here $\theta_1=\theta_2=\theta_0$, where $\cos\theta_0=B_z/B_{\textsc{sf}}$ with $B_{\textsc{sf}}=(8 J_\textsc{afm}+2K)/\mu_s$ being the field of spin-flip. While the out-of-plane components of both sublattices are the same $m_{1z}^0=m_{2z}^0$, the in-plane components are opposite: $\phi_1=\phi_0$ and $\phi_2=\phi_0+\pi$, where $\phi_0$ is arbitrary. Here, we neglect the higher-order in-plane anisotropies present in NiF$_2$~\cite{Moriya60b}, which lead to the preferable values for $\phi_\nu$, as well as to the small in-plane canting~\cite{Moriya60b}.


\section{Magnon dispersion relation}\label{sec:disp}

\begin{figure}
	\includegraphics[width=\columnwidth]{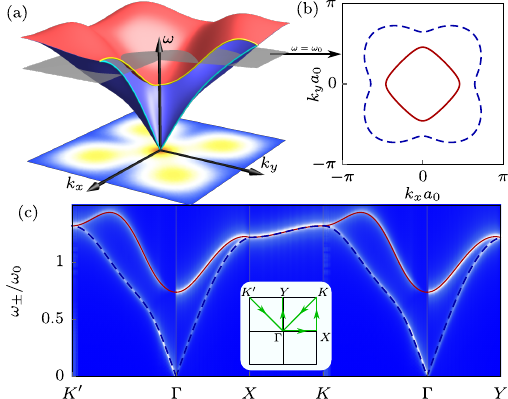}
	\caption{Magnon spectra~\eqref{eq:omega-main} for the easy planar NiF$_2$. (a)~Shows the evolution of two branches $\omega_+$ (red) and $\omega_-$ (blue) within the 1st Brillouin zone. The value of splitting $\Delta\omega_{\vec{k}} = \omega_+-\omega_-$ is shown by the color coding at the bottom. Surface of the constant energy $\omega=$const is shown on panel~(b). (c)~Shows the comparison of analytical prediction~\eqref{eq:omega-main} and results of numerical simulations along the path $K'$--$\Gamma$--$X$--$K$--$\Gamma$--$Y$. Here we use the following parameters: $\varepsilon=0.25$, $\eta=0.1$, $\kappa=0.5$, and $B_z= J_\textsc{afm}/\mu_s$.} \label{fig:spec}
\end{figure}

Here we consider a model, in which the dynamics of each magnetic moment $\vec{m}_\nu$  is described by the classical Landau-Lifshitz equation 
\begin{equation}\label{eq:LL}
	\dot{\vec{m}}_{\nu}(\vec{r}_{\vec{n}})=\frac{\gamma}{\mu_s}\left[\vec{m}_{\nu}(\vec{r}_{\vec{n}})\times\frac{\partial\mathcal{H}}{\partial\vec{m}_{\nu}(\vec{r}_{\vec{n}})}\right],
\end{equation}
where $\gamma=|g|\mu_{\textsc{b}}/\hbar>0$ is the electron gyromagnetic ratio; $\vec{r}_{\vec{n}}=\vec{R}_{\vec{n}}$, and $\vec{r}_{\vec{n}}=\vec{R}'_{\vec{n}}$ for the sublattices `1' and `2', respectively. The infinite set of equations \eqref{eq:LL} coupled via Hamiltonian $\mathcal{H}$ describes the collective dynamics of all magnetic moments. 
To describe spin waves, it is convenient to proceed from the dynamics of the moments $\vec{m}_\nu$ to their deviations from the ground state $\vec{m}^0_\nu=(-1)^{\nu-1}\sin\theta_0(\vec{e}_x\cos\phi_0+\vec{e}_y\sin\phi_0)+\cos\theta_0\vec{e}_z$. To this end, we utilize the classical form of the Holstein-Primakoff representation~\cite{Tyablikov75,Holstein40}

\begin{equation}\label{eq:Tyabl}
 \vec{m}_\nu=\vec{m}_\nu^0\left(1-|\psi_\nu|^2\right)+\sqrt{2-|\psi_\nu|^2}(\vec{e}^+_\nu\psi_\nu+\vec{e}^-_\nu\psi^*_\nu).
\end{equation}
Here, the creation and annihilation operators are replaced by a complex-valued function $\psi_\nu$ encoding the deviation of $\vec{m}_\nu$ from its equilibrium direction $\vec{m}^0_\nu$. The form of the basis vectors  $\vec{e}^\pm_\nu=\frac{1}{2}(\cos\theta_\nu\cos\phi_\nu\mp i\sin\phi_\nu)\vec{e}_x+\frac{1}{2}(\cos\theta_\nu\sin\phi_\nu\pm i\cos\phi_\nu)\vec{e}_y-\frac12\sin\theta_\nu\vec{e}_z$ guarantees fulfillment of the constraint $|\vec{m}_\nu|=1$~\footnote{One should take into account the following properties $\vec{m}_\nu^0\cdot\vec{e}^\pm_\nu=0$, $\vec{e}^\pm_\nu\cdot\vec{e}^\pm_\nu=0$, and $\vec{e}^\pm_\nu\cdot\vec{e}^\mp_\nu=1/2$.}. In terms of $\psi_\nu$-functions, the Landau-Lifshitz equations~\eqref{eq:LL} take the Schr{\"o}dinger-like form
\begin{equation}\label{eq:LL-psi}
i\dot{\psi}_\nu(\vec{r}_{\vec{n}})=\frac{\gamma}{\mu_s}\frac{\partial\mathcal{H}}{\partial\psi^*_\nu(\vec{r}_{\vec{n}})}.
\end{equation}
Linearization of equations \eqref{eq:LL-psi} on a periodic lattice with subsequent transition to $k$-space allows one to proceed from an infinite set of equations \eqref{eq:LL-psi} to a single Schr{\"o}dinger equation formulated for a 4-component Nambu spinor, see Appendix~\ref{app:model}. The obtained in this way magnon dispersion relation is as follows
\begin{align}\label{eq:omega-main}
\nonumber&\omega_\pm(\vec{k}) = \omega_0 \sqrt{ \mathcal{F}^2_{\vec{k}} - \mathcal{B}^2 +\Lambda^2_{\textsc{alt}}(\vec{k})+ {\Omega_{\vec{k}}^\text{c}}^2 - {\Omega_{\vec{k}}^\text{s}}^2 \pm 2\Omega^0_{\vec{k}}},\\
&\Omega^0_{\vec{k}}=\sqrt{\left[\mathcal{F}_{\vec{k}}\Omega_{\vec{k}}^\text{c}+\mathcal{B}\Omega_{\vec{k}}^\text{s}\right]^2+\Lambda^2_{\textsc{alt}}(\vec{k})\left[\mathcal{F}^2_{\vec{k}}
-{\Omega_{\vec{k}}^{{\text{s}}}}^2\right]}.    
\end{align}
Here $\omega_0=4\gamma J_{\textsc{afm}}/\mu_s$ is the characteristic frequency of the system. We have also introduced the following notations $\mathcal{F}_{\vec{k}} = 1 + \mathcal{B}+ \eta\Omega_\textsc{fm}(\vec{k})$, $\mathcal{B} = \frac{\kappa}{4}\sin^2\theta_0$ with $\kappa=K/J_{\textsc{afm}}$ being the dimensionless anisotropy constant. $\eta=J_\textsc{fm}/J_{\textsc{afm}}$ is the dimensionless ferromagnetic exchange and $\Omega_\textsc{fm}(\vec{k}) = \sin^2\frac{k_xa_0}{2} + \sin^2\frac{k_ya_0}{2}$. The effect of A$\ell$M in encoded in $\Lambda_{\textsc{alt}}(\vec{k})=\varepsilon\Omega_{\textsc{alt}}(\vec{k})$, where $\varepsilon = J_{\textsc{alt}}/J_{\textsc{afm}}$ determines the order of magnitude of the altermagnetic effects and $\Omega_\textsc{alt}(\vec{k}) = \sin(k_xa_0)\sin(k_ya_0)$. We also introduce $\Omega_{\vec{k}}^{\text{c}}=\cos^2\theta_0\Omega_{xy}(\vec{k})$ and $\Omega_{\vec{k}}^{\text{s}}=\sin^2\theta_0\Omega_{xy}(\vec{k})$ with  $\Omega_{xy}(\vec{k})=\cos\frac{k_xa_0}{2}\cos\frac{k_ya_0}{2}$. The applied magnetic field $B_z$ enters dispersion relations \eqref{eq:omega-main} through the canting angle $\theta_0$.


\begin{figure}
	\includegraphics[width=\columnwidth]{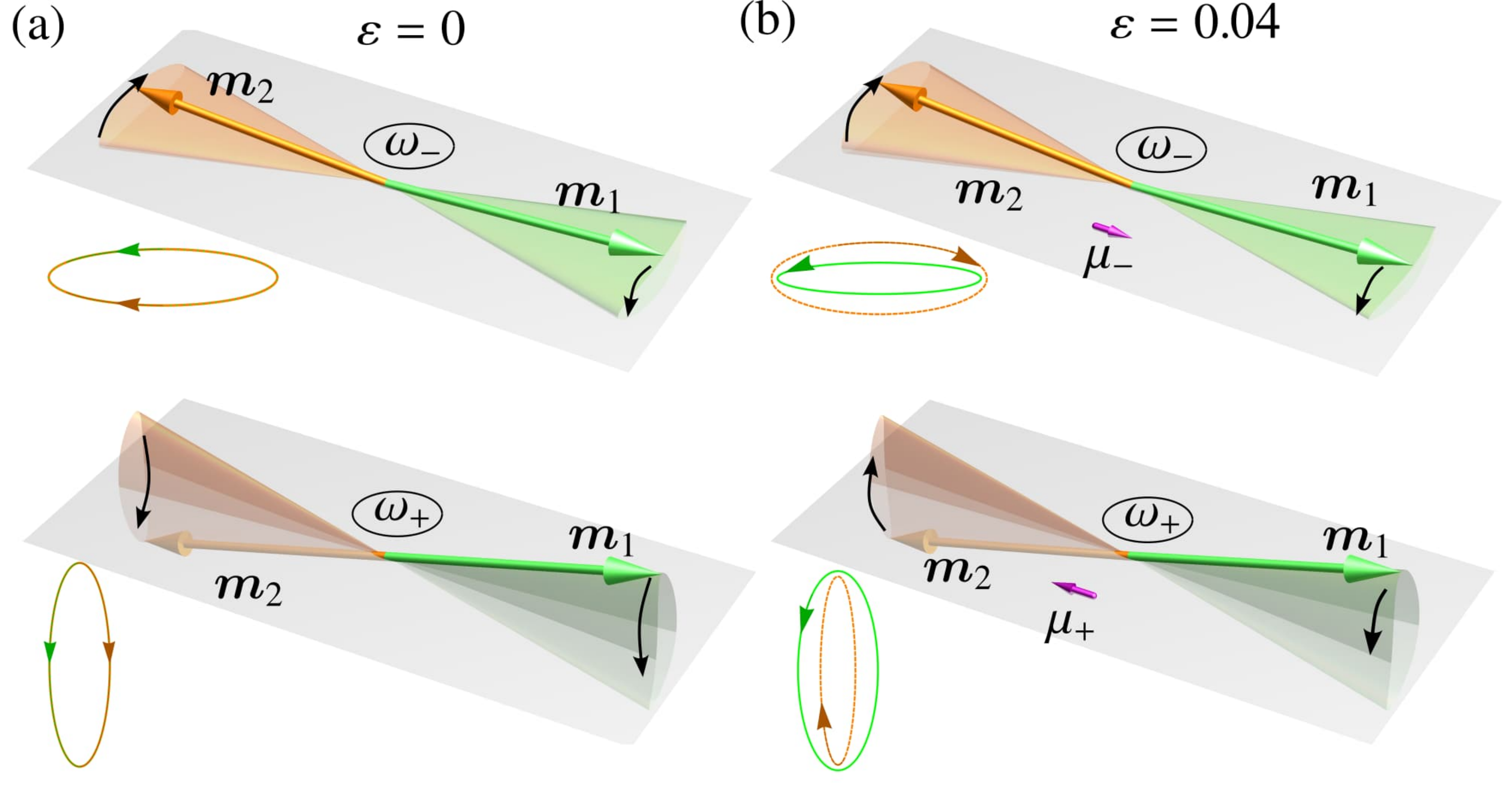}
	\caption{ 
    Precession of the neighboring magnetic moments belonging to different sublattices in the spin wave with the wave vector $(k_x a_0, k_y a_0) = (0.5, 0.5)$. The top and bottom rows correspond to the gapless $\omega_-$ and gapped $\omega_+$ modes, respectively. Panels (a) and (b) show the antiferromagnetic ($\varepsilon=0$) and altermagnetic ($\varepsilon=0.04$) cases, respectively. In the latter case, the magnon magnetic moments $\vec{\mu}_\pm$ are generated. In both cases, the magnetic field is absent ($\theta_0=\pi/2$), $\kappa=0.1$, and $\eta=0.5$. The single-spin dynamics was reconstructed using Eq.~\eqref{eq:dm-sw}. The precession ellipses and the precession directions are shown from the perspective of an observer located at the tip of the vector $\vec{m}^0_1$. The easy plane is indicated in gray.}
  \label{fig:singleDyn}
\end{figure}

The calculated magnon spectrum is shown in Fig.~\ref{fig:spec}. It consists of two branches that differ in the polarizations of the spin waves. In both branches, magnetic moments $\vec{m}_\nu$ demonstrate elliptical precession above their equilibrium directions $\vec{m}^0_\nu$. However, the orientations of the precession ellipses are different in different branches, see Fig.~\ref{fig:singleDyn}. While in the upper gapped branch, the major semi-axis is perpendicular to the easy plane, in the lower gapless branch, it is parallel to the easy plane~\footnote{This is correct for $B_z=0$. In the magnetic field, the planes of the precession ellipses rotate according to the canting.}. As the $\Gamma$-point is approached ($\vec{k}\to\vec
0$), the polarization of the lower mode becomes linear. I.e., the dynamics of the magnetic moments reduces to a uniform rotation of the magnetizations of the two sublattices within the easy plane. Since this type of motion does not cost energy, the lower mode is gapless at the $\Gamma$-point. This is consistent with the Goldstone theorem, because the magnetic ordering spontaneously breaks the continuous SO(2) symmetry of the Hamiltonian related to the arbitrariness of $\phi_0$. Since the precession of magnetic moments upon excitation of the upper mode is predominantly perpendicular to the easy plane, the energy of these excitations does not vanish even at the $\Gamma$-point. The corresponding gap is $\Delta=\omega_0\sqrt{\kappa\sin^2\theta_0+4\cos^2\theta_0}$.

Thus, due to the anisotropy and/or magnetic field, two magnon branches are not degenerate in the antiferromagnetic limit ($\varepsilon=0$). A$\ell$M ($\varepsilon \neq 0$) enhances mode splitting along the diagonal $\Gamma-K$ and $\Gamma-K'$ directions; however, it does not result in the alternating-sign splitting characteristic of easy-axis A$\ell$Ms~\cite{Gomonay24a,Yershov24b,Kravchuk25a}. The analogous situation was previously considered for the easy-planar MnTe~\cite{Kravchuk25a}, and it was shown that the altermagnetic nature of the magnon spectrum manifests itself in terms of the spin splitting parameter $\lambda$, which comprises both the magnon eigenenergy (eigenfrequency) and the magnon magnetic moment. In easy-planar A$\ell$Ms, the latter is generated by A$\ell$M. In detail, this phenomenon is discussed in the next section. 

Finally, we would like to emphasize that the analytical expression \eqref{eq:omega-main} perfectly agrees with the magnon dispersion obtained from numerical spin-lattice simulations, see Appendix~\ref{app:simuls} for the details.

\section{Magnon Magnetic Moment}

Let us now compute the magnetic moment carried by one magnon~\cite{Ashcroft76,Neumann20,Yershov24b}:
\begin{equation}\label{eq:mu}
    \vec{\mu}_\pm = -\hbar\, \partial_{\vec{B}} \omega_\pm .
\end{equation}
The form of $\vec{\mu}_\pm$ essentially depends on the type of the magnetocrystalline anisotropy and direction of the applied field. In the easy-axial crystals in vanishing magnetic field, $\vec{\mu}_\pm$ is collinear to the ground-state N{\'e}el vector $\vec{n}_0=(\vec{m}^0_1-\vec{m}^0_2)/2$, which is parallel to the easy-axis. We observe that the value of $\vec{\mu}_{\pm}$ is $\vec{k}$-independent, and its direction is opposite for the two magnon branches. A finite magnetic field $B_z$ applied along the easy axis does not affect the value of $\vec{\mu}_\pm$~\footnote{We assume that the applied magnetic field is smaller than the spin-flop field.}. The latter is a direct consequence of the linearity of the magnon eigenfrequencies with respect to $B_z$: both magnon branches acquire the Larmor frequency shift, with opposite signs for the two branches. The behavior of $\vec{\mu}_\pm$ is identical in conventional AFMs and A$\ell$Ms, i.e., A$\ell$M does not affect the magnon magnetic moment in easy-axial systems~\footnote{If the magnetic field is applied along the anisotropy axis or absent.}. The latter result was obtained for the easy-axial $d$-wave~\cite{Yershov24b} and $g$-wave~\cite{Kravchuk25a} A$\ell$Ms.

\begin{figure}
	\includegraphics[width=\columnwidth]{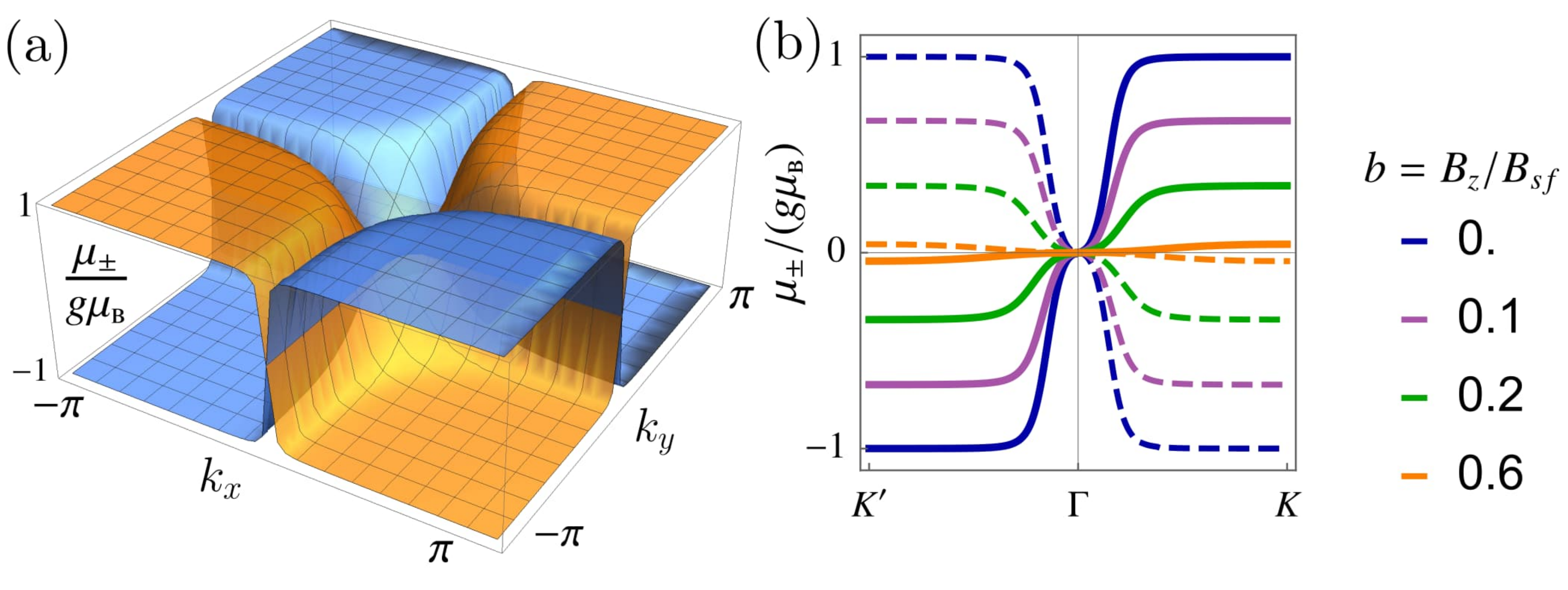}
	\caption{
    Distribution of the magnon magnetic moment determined by Eq.~\eqref{eq:mu-rutile} within the 1st Brillouin zone. 
    Panel (a) shows the case $B_z=0$. Magnetic moments, which correspond to the upper ($\mu_+^{\text{gs}}$) and lower  ($\mu_-^{\text{gs}}$) magnon branches, are shown by blue and yellow surfaces, respectively.  
    Panel (b) shows the effect of suppression of $\mu_\pm^{\text{gs}}$ by the applied magnetic field. The $k$-dependencies are shown along the path $K' - \Gamma - K$ by solid and dashed lines, which correspond to the lower and upper branches, respectively. 
    The parameters used in the calculations are $\kappa=0.08$, $\varepsilon=0.1$, and $\eta=0.5$. }
    \label{fig:magnon_moment} 
\end{figure}

In the easy-planar systems, the behavior of $\vec{\mu}_\pm$ differs markedly. For vanishing magnetic field, the out-of-plane component $\mu_\pm^z=0$ for both allermagnets and antifferomagnets. The latter is a direct consequence of the quadratic dependence $\omega_\pm=\omega_\pm(B_z^2)$ of the eigenfrequencies on the magnetic field applied parallel to the hard axis. This is evident from equation \eqref{eq:omega-main}, in which $B_z$ enters via the angle $\theta_0$, e.g., for vanishing field $\theta_0=\pi/2$. In contrast to AFMs, however, magnons in easy-planar aletrmagnats gain a finite in-plane magnetic moment $\vec{\mu}_\pm\parallel\vec{n}_0$. Note that the N{\'e}el vector $\vec{n}_0$ is oriented within the easy plane. This is a purely altermagnetic effect; previously, it was predicted for MnTe~\cite{Kravchuk25a}, which is an easy-planar $g$-wave altermagnet. Here, applying a similar technique~\footnote{The technique is based on the application of a small auxiliary $B_x$ parallel to $\vec{n}_0$. Then we use formula \eqref{eq:mu} in the limit $B_x\to0$ keeping $B_z$ finite.} we obtain the component  
\begin{align} \label{eq:mu-rutile}
\nonumber\mu^{\text{gs}}_{\pm}(\vec{k})=&-|g|\mu_{\textsc{b}}\frac{\Lambda_{\textsc{alt}}(\vec{k})\kappa}{4\omega_{\pm}(\vec{k})/\omega_0}\frac{1+\cos^2\theta_0}{\mathcal{B}+\cos^2\theta_0}\sin\theta_0\left(1\pm\frac{\mathcal{G}_{\vec{k}}}{\Omega^0_{\vec{k}}}\right),\\
&\mathcal{G}_{\vec{k}}=\mathcal{F}_{\vec{k}}^2-\sin^2\theta_0\frac{\Omega^2_{xy}(\vec{k})+\frac{\kappa}{4}\mathcal{F}_{\vec{k}}\cos^2\theta_0}{1+\cos^2\theta_0}
\end{align}
of the magnon magnetic moment along the ground-state N{\'e}el vector $\vec{n}_0$ for our $d$-wave model,
for details, see Appendix~\ref{app:mu}. Important that $\mu^{\text{gs}}$ is of altermagnetic origin: it vanishes in the antiferromagnetic limit ($\varepsilon=0$) and possesses the $d$-wave symmetry, i.e. $\mu^{\text{gs}}_\pm(\mathcal{R}_{\pi/2}\vec{k})=-\mu^{\text{gs}}_\pm(\vec{k})$ with $\mathcal{R}$ being the rotation matrix; see Fig.~\ref{fig:magnon_moment}.
The emergence of $\mu^{\text{gs}}_\pm$ in the altermagnetic regime with $\varepsilon\ne0$ is illustrated in Fig.~\ref{fig:singleDyn}(b), which shows that finite $\varepsilon$ results in different amplitudes of the elliptical precession of the magnetic moments $\vec{m}_\nu$ in different sublattices. The latter implies that the contributions of the two sublattices to the magnon magnetic moment are different and do not cancel each other out, as occurs in the antiferromagnetic case~\footnote{Although this simple intuitive explanation of the origin of the magnon magnetic moment works in most cases, there are examples when it is not applicable~\cite{Kravchuk26}.}, see Fig.~\ref{fig:singleDyn}(a). 

Here, we are interested in the influence of $B_z$ on the magnon magnetic moment, as it offers an opportunity to control transport properties by the magnetic field. The latter enters Eq.~\eqref{eq:mu-rutile} via the canting angle $\theta_0$. We find that the magnetic field $B_z$ doesn't affect the symmetry of the distribution of $\mu^{\text{gs}}_\pm(\vec{k})$ over the 1st Brillouin zone; however, it can significantly suppress the amplitude of the magnon magnetic moment, see Fig.~\ref{fig:magnon_moment}(b). This effect occurs because the field-induced canting reduces the magnetization of each sublattice along the direction of the ground state $\vec{n}_0$. Note that in the presence of $B_z$, the out-of-plane component $\mu_\pm^z$ is generated in both the antiferromagnetic and altermagnetic cases. This effect, however, will not be considered here, since it is not of altermagnetic origin.

\begin{figure}	\includegraphics[width=\columnwidth]{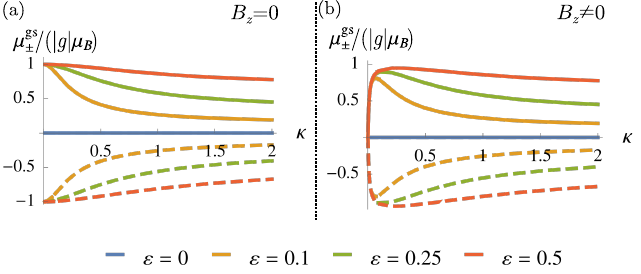}
	\caption{Magnon magnetic moment \eqref{eq:mu-rutile} along the ground state N{\'e}el vector $\vec{n}_0$ as a function of the anisotropy parameter $\kappa$ for various values of the altermagnetic strength $\varepsilon$. Solid and dashed lines represent the lower ($\mu^{\text{gs}}_-$) and upper ($\mu^{\text{gs}}_+$) magnon branches, respectively. Cases with $B_z=0$ and $B_z=0.4 J_\textsc{afm}/\mu_s$  are compared in panels (a) and (b). In all cases, $\eta=0.5$, and $(k_x a_0, k_y a_0) = (\pi/4, \pi/4)$.}
    \label{fig:mu_kappa} 
\end{figure}

\begin{figure}
\includegraphics[width=\columnwidth]{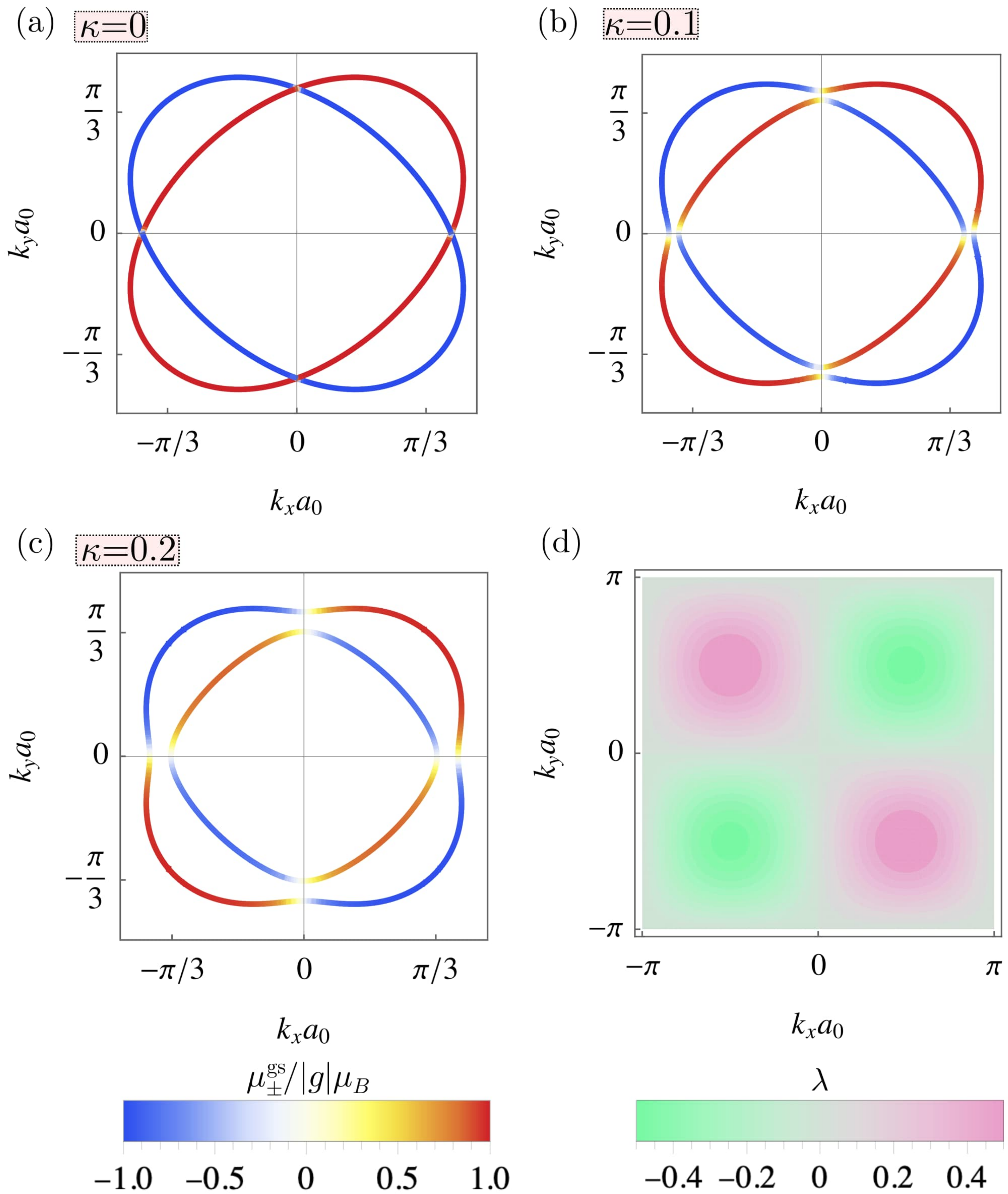}
	\caption{Momentum-space mappings of the magnon magnetic moments $\mu^{\text{gs}}_{\pm}(\vec{k})$ determined in Eq.~\eqref{eq:mu-rutile} along the iso-energy contours $\omega_\pm(\vec{k})=0.85\omega_0$ determined by Eq.~\eqref{eq:omega-main} for various anisotropies $\kappa$ are shown on panels 
    (a)--(c). Panel (d) shows the $d$-wave-symmetry distribution of the spin-splitting parameter $\lambda$ defined in Eq.~\eqref{eq:lambda1} over the first Brillouin zone. Foe all cases, we use the following parameters: $\varepsilon=0.25$, $B_z=0$, $\eta=0.5$. }
  \label{fig:isenMap}
\end{figure}


The crucial role of the altermagnetic parameter, as well as the combined action of the anisotropy and the applied magnetic field on the formation of the magnon magnetic moment $\mu^{\text{gs}}_{\pm}$ along the ground state N{\'e}el vector $\vec{n}_0$, is illustrated in Fig.~\ref{fig:mu_kappa}. In the antiferromagnetic limit ($\varepsilon=0$), one has $\mu^{\text{gs}}_{\pm}=0$ for any values of the anisotropy and the applied field. However, for an arbitrarily small value of $\varepsilon$, a finite $\mu^{\text{gs}}_{\pm}$ is generated. Moreover, if $B_z=0$, then in the limit of the vanishing anisotropy $\kappa\to0$, the magnon magnetic moment $\mu^{\text{gs}}_{\pm}\to\pm|g|\mu_{\textsc{b}}$ reaches values of the magnetic moment of magnons excited over the collinear state of the easy-axial antiferro- and A$\ell$Ms~\cite{Yershov24b,Kravchuk25a}, see Fig.~\ref{fig:mu_kappa}(a). The latter situation is realized for $\kappa<0$. Note that $\mu^{\text{gs}}_{\pm}$ is not defined in the critical case $\kappa=0$ and $B_z=0$; thus, a finite, possibly infinitely small, anisotropy should be present if $B_z=0$. The influence of the finite magnetic field $B_z\ne0$ on the magnetic moment is insignificant for large anisotropies. However, it drastically suppresses $\mu^{\text{gs}}_{\pm}$ in the limit $\kappa\to0$, see Fig.~\ref{fig:mu_kappa}(b). So the control of the magnon magnetic moment by magnetic field is especially effective for small anisotropies.

Finally, we note that although the absolute values of $\mu^{\text{gs}}_{+}$ and $\mu^{\text{gs}}_{-}$ are close, in general case, $\mu^{\text{gs}}_{+}(\vec{k})\ne-\mu^{\text{gs}}_{-}(\vec{k})$.

It is instructive to trace the simultaneous evolution of the magnon dispersion relation $\omega_\pm(\vec{k})$ and the magnon magnetic moment $\mu_\pm^{\text{gs}}(\vec{k})$ as the anisotropy parameter is varied. We do it by means of the isosurfaces of constant energy $\omega_\pm(\vec{k})=\text{const}$ colored according to $\mu_\pm^{\text{gs}}(\vec{k})$ for the case $B_z=0$, see Fig~\ref{fig:isenMap}(a-c). In the non-relativistic limit $\kappa\to0$, we obtain two intersecting, oppositely polarized ellipses rotated by $\pi/2$ relative to each other, see Fig~\ref{fig:isenMap}(a). This is a characteristic feature of $d$-wave A$\ell$Ms. The finite anisotropy opens gaps in the nodal directions, see Fig~\ref{fig:isenMap}(b,c). The latter breaks the $d$-wave symmetry of the direct splitting $\Delta\omega=\omega_+-\omega_-$ of the magnon branches. However, due to the symmetry of the magnon magnetic moment, in particular due to the fact that $\mu_{\pm}^{\text{gs}}=0$ for the nodal directions, one can introduce the universal spin-splitting parameter~\cite{Kravchuk25a}
\begin{equation}\label{eq:lambda}
    \lambda = \frac{1}{\omega_0 |g| \mu_\textsc{b}} \sum_{\nu=\pm} \mu_\nu^{\text{gs}} \omega_\nu,
\end{equation}
which possesses $d$-wave symmetry even beyond the non-relativistic limit. For the case of the obtained magnon dispersion~\eqref{eq:omega-main} and magnetic moment~\eqref{eq:mu-rutile} we obtain
\begin{equation}\label{eq:lambda1}
    \lambda(\vec{k})=-\frac{\Lambda_{\textsc{alt}}(\vec{k})\kappa}{2}\frac{1+\cos^2\theta_0}{\mathcal{B}+\cos^2\theta_0}\sin\theta_0.
\end{equation}
In the limit of the vanishing magnetic field ($\theta_0=\pi/2$), the spin-splitting parameter obtains a simple anisotropy-independent form $\lambda(\vec{k})=-2\Lambda_{\textsc{alt}}(\vec{k})$, which structurally coincides with one previously obtained for $g$-wave A$\ell$Ms~\cite{Kravchuk25a}. The distribution of $\lambda(\vec{k})$ in the first Brillouin zone is universally $d$-wave symmetric, see Fig.~\ref{fig:isenMap}(d).

\section{Thermal spin transport}

\begin{figure}	\includegraphics[width=\columnwidth]{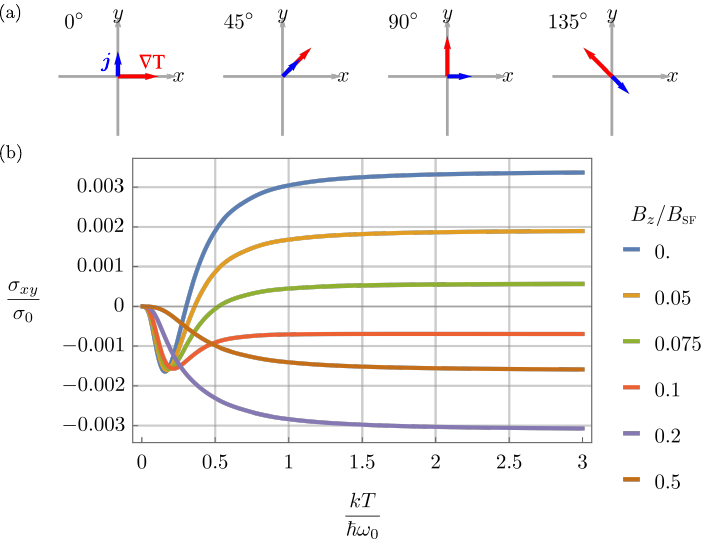}
\caption{(a) Schematic representation of the mutual orientation of the applied temperature gradient $\vec{\nabla} T$ (red arrows) and the induced spin current density $\vec{j}$ (blue arrows) for different orientations ($0^\circ$, $45^\circ$, $90^\circ$, and $135^\circ$)  of the temperature gradient relative to the direction [100]. (b) Temperature dependencies of $\sigma_{xy}/\sigma_0$ for different values of the applied magnetic field $B_z$. The rest of the parameters are $\varepsilon=0.1$, $\kappa=0.08$, $\eta=0.5$.}
  \label{fig:sigmaT}
\end{figure}

As a measurable physical effect directly resulting from the above-described altermagnet-induced magnon magnetic moment, we consider the magnon spin-splitter effect, which is the generation of a flow of magnon magnetic moment in response to an applied temperature gradient. Following standard terminology, we will also refer to this current as a spin current. This effect was theoretically predicted~\cite{Yershov24b,Weissenhofer24,Cui23a}, and recently observed experimentally~\cite{Yang26b} in A$\ell$Ms, where the magnon magnetic moment is not of altermagnetic origin, e.g., in easy-axial $d$-wave A$\ell$Ms~\cite{Yershov24b}.

The transport of magnons in the presence of a temperature gradient $\vec{\nabla}T$ is described within the framework of the semi-classical Boltzmann transport theory. In the relaxation time approximation, the deviation of the magnon distribution function from the equilibrium Bose-Einstein distribution $n^0_{\vec{k},\nu}=[e^{\mathcal{E}_{\vec{k},\nu}/(k_{\textsc{b}}T)}-1]^{-1}$ is given by $\delta n_{\vec{k},\nu} \approx -\tau_{\mathrm{rlx}} (\vec{v}_{\vec{k},\nu} \cdot \vec{\nabla}T) \partial_T n^0_{\vec{k},\nu}$. Here $\tau_{\text{rlx}}$ is the relaxation time, and $\vec{v}_{\vec{k},\nu}=\partial_{\vec{k}}\omega_\nu(\vec{k})$ is the magnon group velocity. The resulting spin current density $\vec{j}$, representing the flow of the magnetic moment component along the ground-state N{\'e}el vector, is determined as:
\begin{equation}
    j_\alpha = \frac{1}{L_x L_y} \sum\limits_{\vec{k},\nu} \mu^{\text{gs}}_{\nu}(\vec{k}) (v_{\vec{k},\nu})_\alpha \delta n_{\vec{k},\nu} = \sigma_{\alpha\beta}\partial_\beta T,
\end{equation}
where $L_x$ and $L_y$ denote the lateral dimensions of the film and indices $\alpha, \beta$ denote the Cartesian coordinates. The thermal spin conductivity tensor $\sigma_{\alpha\beta}$ is then expressed as follows:
\begin{equation}\label{eq:sigma-ab}
    \sigma_{\alpha\beta}=\frac{\tau_{\mathrm{rlx}}}{L_xL_y}\sum\limits_{\vec{k}\in\text{1.BZ}}\sum\limits_{\nu=\pm}c_{\vec{k},\nu}(v_{\vec{k},\nu})_\alpha(v_{\vec{k},\nu})_\beta,
\end{equation}
where $c_{\vec{k},\nu}=-\mu_\nu^{\text{gs}}(\vec{k})\partial_Tn^0_{\vec{k},\nu}$ is the spin capacity per magnon~\cite{Yershov24b}.


Based on \eqref{eq:omega-main} and \eqref{eq:mu-rutile}, we deduce that $\mu_\pm^{\text{gs}}$ is odd with respect to a reflection transformation $k_x\to-k_x$ or $k_y\to-k_y$, while $\omega_\pm$ is even. Consequently, $(v_{\vec{k},\nu})_x$ is odd and even under the reflection of $k_x$ and $k_y$, respectively; and it is vice versa for $(v_{\vec{k},\nu})_y$. Using these symmetries, from \eqref{eq:sigma-ab}, we conclude that $\sigma_{xx}=\sigma_{yy}=0$. For non-zero out-of-diagonal components $\sigma_{xy}=\sigma_{yx}$, the mutual orientation of the spin current $\vec{j}$ and the temperature gradient $\vec{\nabla}T=|\vec{\nabla}T|(\cos\chi\vec{e}_x+\sin\chi\vec{e}_y)$ is determined by the formula $\vec{j}=\sigma_{xy}|\vec{\nabla}T|(\sin\chi\vec{e}_x+\cos\chi\vec{e}_y)$. Thus, the spin current is perpendicular to the temperature gradient if the latter is applied in a nodal direction [100] or [010], see Fig.~\ref{fig:sigmaT}(a). The situation $\vec{j}\parallel\vec{\nabla}T$ takes place if the temperature gradient is applied in a direction of the maximal altermagnetic splitting [110] or [$\bar{1}10$], see Fig.~\ref{fig:sigmaT}(a). The same mutual orientations of $\vec{j}$ and $\vec{\nabla}T$ were previously predicted for easy-axial rutiles~\cite{Yershov24b}. Note that the structure discussed above of tensor $\sigma_{\alpha\beta}$ corresponds to the orientation of the reference frame $x0y$ along the nodal directions, see Fig.~\ref{fig:model}. In the rotated reference frame, the components of tensor $\sigma_{\alpha\beta}$ should be transformed accordingly.

The out-of-diagonal component can be presented in form $\sigma_{xy}=\sigma_0\varsigma(k_{\textsc{b}}T/(\hbar\omega_0))$, where $\sigma_0=\tau_{\text{rlx}}\omega_0|g|\mu_{\textsc{b}}/\hbar$ and the dimensionless function $\varsigma(k_{\textsc{b}}T/(\hbar\omega_0))$ is defined in Eq.~\eqref{eq:sigma-dmnls} and shown in Fig.~\ref{fig:sigmaT}(b) for different values of magnetic field. One can see that the temperature dependence of $\sigma_{xy}$ can be essentially non-monotonous in the regime when $\varepsilon$ is comparable to $\kappa$. In the small temperature limit $k_{\textsc{b}}T\ll\Delta$, the dominant contribution to the spin conductivity comes from the lower gapless branch. In this limit
\begin{equation}\label{eq:sigma-low-T}
    \sigma_{xy}\approx-\sigma_0\frac{3\zeta(3)}{8\pi}\frac{\varepsilon\kappa}{1+2\eta}\frac{\sin\theta_0}{\mathcal{B}+\cos^2\theta_0}\left(\frac{k_{\textsc{b}}T}{\hbar\omega_0}\right)^2,
\end{equation}
where $\zeta(x)$ is Riemann zeta function. The sign of the prefactor in the quadratic temperature dependence \eqref{eq:sigma-low-T}  is solely determined by the sign of the altermagnetic parameter $\varepsilon$, and is not influenced by the magnetic field, which enters Eq.~\eqref{eq:sigma-low-T} via $\theta_0$. As the temperature increases, the upper gapped branch becomes populated. The magnons of this branch possess a magnetic moment of the opposite sign to that of the magnons in the lower gapless branch, see Figs.~\ref{fig:magnon_moment},~\ref{fig:mu_kappa}, and partially compensate the contribution of the lower branch to the spin conductivity. As a result, the temperature dependence of $\sigma_{xy}$ can possess an extremum and can change sign with temperature, see Fig.~\ref{fig:sigmaT}(b). Thus, in the high-temperature regime $k_{\textsc{b}}T\gg\Delta$, the value of $\sigma_{xy}$ is determined by a delicate balance between contributions of the two branches. In this regime, the influence of the applied magnetic field is essentially strong. This is because, as the field increases, the upper magnon branch becomes increasingly flat, and consequently, its contribution to transport properties decreases. The strong influence of the magnetic field on $\sigma_{xy}$ is demonstrated in Fig.~\ref{fig:sigmaT}(b).

\begin{figure}
\centering
\includegraphics[width=\columnwidth]{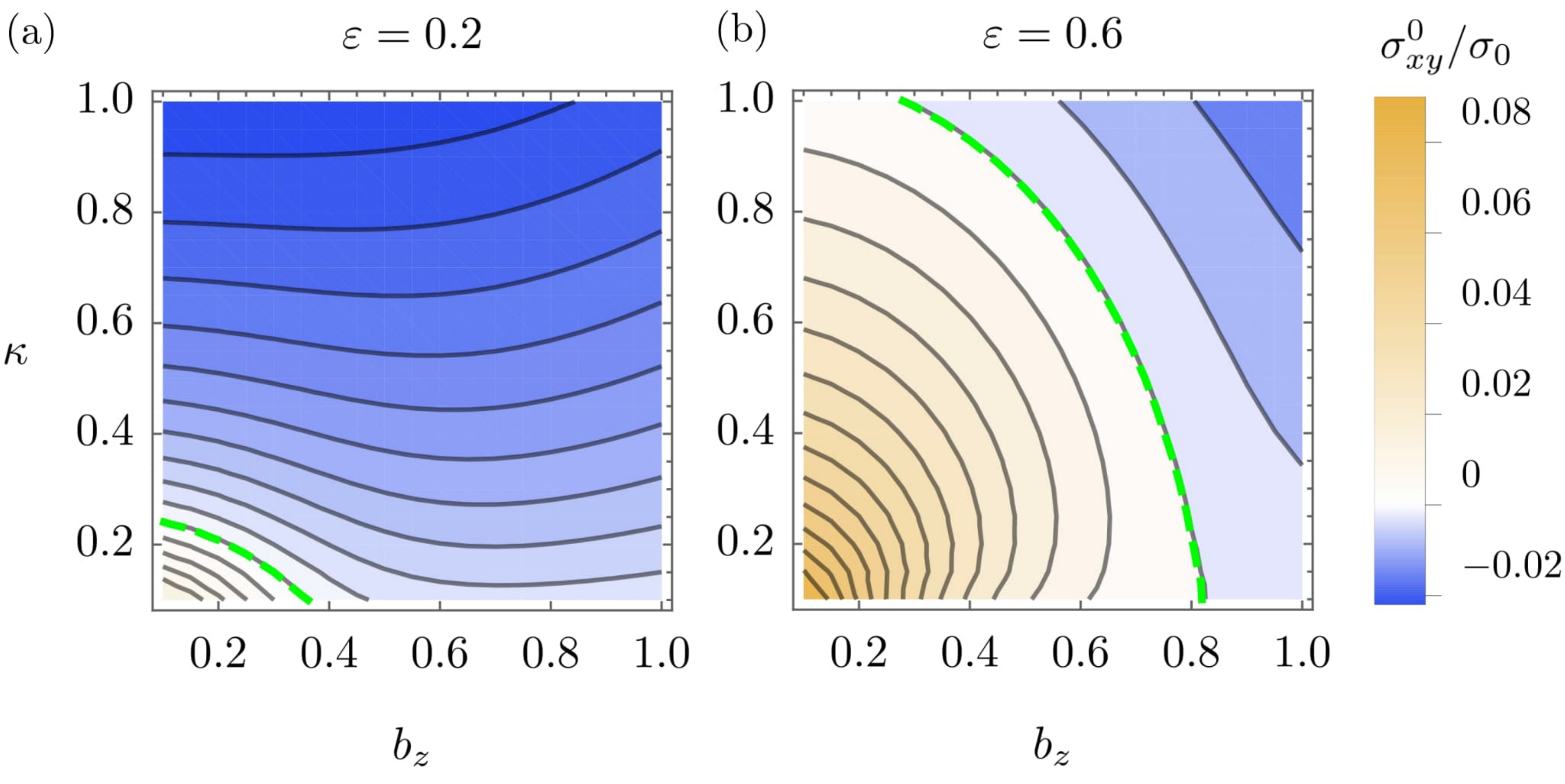}
	\caption{Map of the high-temperature asymptotic value $\sigma_{xy}^0$ of the spin conductivity determined in Eq.~\eqref{eq:sigma-xy-HT} as a function of the magnetic field $b_z= B_z\mu_s / (4J_\textsc{AFM})$ and the anisotropy coefficient $\kappa$ for various values of $\varepsilon$. Here $\eta=0.5$. The isoline $\sigma_{xy}^0=0$ is shown in green.}
  \label{fig:mapsigma}
\end{figure}

In the high-temperature regime, $k_{\textsc{b}}T\gg\hbar\omega_0$, the spin conductivity $\sigma_{xy}$ loses its temperature dependence, approaching horizontal asymptotics $\sigma_{xy}^0$ determined by Eq.~\eqref{eq:sigma-xy-HT}. In this regime, as discussed above, the system is very sensitive to the applied magnetic field, and the latter can be used for effective control of the spin conductivity, including flipping its sign. Fig.~\ref{fig:mapsigma} demonstrates that the efficiency of the spin conductivity control by the magnetic field increases as $\varepsilon$ increases, i.e., for larger $\varepsilon$, the range of $\sigma_{xy}^0$, which can be tuned by the magnetic field is larger.

\section{Summary and outlook} 

Here we demonstrate that easy-planar $d$-wave A$\ell$Ms provide a convenient platform for the generation and manipulation of spin currents driven by magnons. This is because of two reasons: (i) A$\ell$M leads to the generation of the $k$-dependent magnetic moment for magnons, and (ii) the magnon-driven spin currents generated by the temperature gradient (spin-splitter effect) can be efficiently controlled by the external magnetic field. Without a magnetic field, the value of the spin conductivity is determined by a delicate balance between the opposite-sign contributions from two magnon branches (if the temperature is not too low): the lower gapless and upper gapped ones. The applied magnetic field flattens the upper branch, reducing its contribution to the transport properties, and shifting the balance in favor of the lower branch's contribution. This opens the possibility of effectively controlling magnon spin currents with a magnetic field.

Conducting this study, we found that in the presence of A$\ell$M, the magnetic field significantly influences magnon polarization. Specifically, we have also observed that the application of a magnetic field may reverse the direction of precession of one of the sublattices. In future work, we will explore this phenomenon in more detail in view of potential applications in magnonic devices.

\section*{Acknowledgment}
We thank Ulrike Nitzsche for the technical support. This work was supported by the German Federal Ministry of Research, Technology and Space (BMFTR) through the GU-QuMat project (01DK24008) and by  Deutsche Forschungsgemeinschaft (DFG, German Research Foundation) through the Sonderforschungsbereich SFB 1143, grant No. YE 232/2-1, and through the W{\"u}rzburg-Dresden Cluster of Excellence ctd.qmat – Complexity, Topology and Dynamics
in Quantum Matter (EXC 2147, project-id 390858490).

\appendix

\section{Model and spin waves}\label{app:model}

Here we present the explicit form of the exchange part of the magnetic Hamiltonian discussed in Section~\ref{sec:model}. The dominant antiferromagnetic interaction~($J_{\textsc{afm}}>0$) between sublattices has the following explicit form
\begin{equation}\label{eq:app-Hafm-real}
	\mathcal{H}_{\textsc{afm}}= J_{\textsc{afm}}\sum\limits_{\langle\vec{R}_{\vec{n}},\vec{R}'_{\vec{n}}\rangle}\vec{m}_1(\vec{R}_{\vec{n}})\cdot\vec{m}_2(\vec{R}'_{\vec{n}}),
\end{equation}
where $\vec{R}'_{\vec{n}}$ counts the nearest neighbors of $\vec{R}_{\vec{n}}$. Hamiltonian
\begin{equation}\label{eq:app-Halt-real}
\begin{split}
    \mathcal{H}_{\textsc{alt}}=\frac{J_{\textsc{alt}}}{2}\sum\limits_{\vec{R}_{\vec{n}}}\Biggl\{&\sum\limits_{\sigma\in \left\{\nwarrow,\searrow\right\}}\left[\vec{m}_1\left(\vec{R}_{\vec{n}}\right)\cdot \vec{m}_1\left(\vec{R}_{\vec{n}}+2\delta\vec{R}_{\sigma}\right)\right.\\
    -&\vec{m}_2\left(\vec{R}_{\vec{n}}\right)\cdot \vec{m}_2\left(\vec{R}_{\vec{n}}+2\delta\vec{R}_{\sigma}\right)] \\
    -&\sum\limits_{\sigma\in \left\{\nearrow,\swarrow\right\}}\left[\vec{m}_1\left(\vec{R}_{\vec{n}}\right)\cdot \vec{m}_1\left(\vec{R}_{\vec{n}}+2\delta\vec{R}_{\sigma}\right)\right.\\
   -&\vec{m}_2\left(\vec{R}_{\vec{n}}\right)\cdot \vec{m}_2\left(\vec{R}_{\vec{n}}+2\delta\vec{R}_{\sigma}\right)]\Biggr\},
\end{split}
\end{equation}
models the altermagnetic properties with $|J_{\textsc{alt}}|\ll J_{\textsc{afm}}$, where $\vec{\delta R}_\nearrow=-\vec{\delta R}_\swarrow=\frac{a_0}{2}(\vec{e}_x+\vec{e}_y)$ and $\vec{\delta R}_\nwarrow=-\vec{\delta R}_\searrow=\frac{a_0}{2}(-\vec{e}_x+\vec{e}_y)$ The intra-sublatte ferromagnetic exchange coupling is as follows
\begin{equation}\label{eq:app-Hfm-real}
\begin{split}
	\mathcal{H}_{\textsc{fm}}= -\frac{J_{\textsc{fm}}}{2}\sum\limits_{\vec{R}_{\vec{n}}}\sum\limits_{\sigma}\bigl[&\vec{m}_1(\vec{R}_{\vec{n}})\cdot\vec{m}_1(\vec{R}_{\vec{n}}+\delta\vec{R}_{\sigma})\\
	+&\vec{m}_2(\vec{R}_{\vec{n}})\cdot\vec{m}_2(\vec{R}_{\vec{n}}+\delta\vec{R}_{\sigma})\bigr],
\end{split}
\end{equation}
where $J_{\textsc{fm}}>0$, $\sigma\in\{\rightarrow,\leftarrow,\uparrow,\downarrow\}$ and displacement vectors $\delta\vec{R}_{\rightarrow}=-\delta\vec{R}_{\leftarrow}=a_0\vec{e}_x$, $\delta\vec{R}_{\uparrow}=-\delta\vec{R}_{\downarrow}=a_0\vec{e}_y$ runs over the four intra-layer neighbors. The anisotropy $\mathcal{H}_{\textsc{an}}$ and Zeeman $\mathcal{H}_{\textsc{z}}$ components are given in the main text.

Here we utilize the Fourier transforms on the periodic lattice 
\begin{equation}\label{eq:FT-discr}
	\begin{split}
		&\psi_\nu(\vec{R}_{\vec{n}})=\frac{1}{\sqrt{N}}\sum\limits_{\vec{k}\in1.\text{BZ}}\hat{\psi}_\nu(\vec{k})e^{i\vec{k}\cdot\vec{R}_{\vec{n}}},\\
		&\hat{\psi}_\nu(\vec{k})=\frac{1}{\sqrt{N}}\sum\limits_{\vec{R}_{\vec{n}}}\psi_\nu(\vec{R}_{\vec{n}})e^{-i\vec{k}\cdot\vec{R}_{\vec{n}}}	
	\end{split}
\end{equation}
supplemented with the completeness relation $\sum_{\vec{R}_{\vec{n}}}e^{i(\vec{k}-\vec{k}')\cdot \vec{R}_{\vec{n}}}=N\delta_{\vec{k},\vec{k}'}$. Here $N$ is the number of magnetic magnetic unit cells. Applying the Fourier transform~\eqref{eq:FT-discr} to linearized Eq.~\eqref{eq:LL-psi}, we obtain
\begin{equation}\label{eq:LL-psi-k}
    i\dot{\hat{\psi}}_\nu(\vec{k})=\frac{\gamma}{\mu_s}\frac{\partial\mathcal{H}^{(2)}}{\partial\hat{\psi}^*_\nu(\vec{k})},
\end{equation}
where $\mathcal{H}^{(2)}=\mathcal{H}_{\textsc{afm}}^{(2)}+\mathcal{H}_{\textsc{fm}}^{(2)}+\mathcal{H}_{\textsc{alt}}^{(2)}+\mathcal{H}_{\textsc{an}}^{(2)}+\mathcal{H}_{\textsc{z}}^{(2)}$ is harmonic (with respect to $\hat{\psi}_\nu$) part of Hamiltonian $\mathcal{H}$. 

Substituting the classical Holstein-Primakoff-Tyablikov representation~\eqref{eq:Tyabl} into the total Hamiltonian, expanding it up to the second order with respect to $\psi_\nu$, and applying the Fourier transform~\eqref{eq:FT-discr}, we derive individual harmonic contributions in $k$-space:
\begin{equation}\label{eq:Hz-k-app}
\mathcal{H}_{\textsc{z}}^{(2)}\!=\!4J_\textsc{afm}b_z\cos\theta_0\!\sum\limits_{\vec{k}\in\text{1.BZ}}\!\!\left[|\hat{\psi}_1(\vec{k})|^2+|\hat{\psi}_2(\vec{k})|^2\right]\!,
\end{equation}
\begin{equation}\label{eq:Halt-k-app}
\mathcal{H}_{\textsc{alt}}^{(2)}\!=\!4J_{\textsc{afm}}\!\sum\limits_{\vec{k}\in\text{1.BZ}}\!\!\varepsilon\,\Omega_\textsc{alt}(\vec{k})\!\left[|\hat{\psi}_1(\vec{k})|^2-|\hat{\psi}_2(\vec{k})|^2\right]\!,
	\end{equation}
 \begin{equation}\label{eq:Hfm-k-app}
\mathcal{H}_{\textsc{fm}}^{(2)}\!=\!4J_{\textsc{afm}}\!\sum\limits_{\vec{k}\in\text{1.BZ}}\!\!\eta\,\Omega_\textsc{fm}(\vec{k})\!\left[|\hat{\psi}_1(\vec{k})|^2+|\hat{\psi}_2(\vec{k})|^2\right]\!,
\end{equation}
\begin{widetext}
\begin{equation}\label{eq:Hafm-k-app}
\mathcal{H}_{\textsc{afm}}^{(2)}=4J_{\textsc{afm}}\!\!\sum\limits_{\vec{k}\in\text{1.BZ}}\!\!\left\{-\frac{\cos 2\theta_0}{2}\left[|\hat{\psi}_1(\vec{k})|^2+|\hat{\psi}_2(\vec{k})|^2\right]+\left[\Omega^{s}_{xy}(\vec{k})\hat{\psi}_1(\vec{k})\hat{\psi}_2(-\vec{k})-\Omega^{c}_{xy}(\vec{k})\hat{\psi}_1(\vec{k})\hat{\psi}^*_2(\vec{k})\right]+\text{c.c.}\right\},
\end{equation}

\begin{equation}\label{eq:Han-k-app}
	\mathcal{H}_{\textsc{an}}^{(2)}=\frac{J_\textsc{afm}}{2}\sum\limits_{\nu=1,2}\sum\limits_{\vec{k}\in\text{1.BZ}}\kappa\left[(1-3\cos^2\theta_0)|\hat{\psi}_\nu(\vec{k})|^2+\sin^2\theta_0 \hat{\psi}_\nu(\vec{k})\hat{\psi}_\nu(-\vec{k})+\text{c.c.}\right],
\end{equation}
\end{widetext}
where $b_z = B_z \mu_s/(4J_\textsc{afm})$ and the rest of the notations are introduced in Section~\ref{sec:disp}.

With total harmonic Hamiltonian $\mathcal{H}^{(2)}$, the couple of equations \eqref{eq:LL-psi-k} together with their complex conjugated counterparts form a closed set of four equations
\begin{equation}\label{eq:Psi-app}
	i\dot{\hat{\vec{\Psi}}}_{\vec{k}}=\hat{\eta}\hat{\mathbb{H}}_{\vec{k}}\hat{\vec{\Psi}}_{\vec{k}},
\end{equation}

\begin{equation}\label{eq:matrix-H-app}
	\nonumber \frac{\hat{\mathbb{H}}_{\vec{k}}}{\omega_0}\!=\!\!\begin{bmatrix}
		\mathcal{F}_{ \vec{k}}+\Lambda_{\textsc{alt}} & -\Omega_{xy}^\text{c}(\vec{k}) & \mathcal{B}  & \Omega_{xy}^\text{s}(\vec{k})  \\
		-\Omega_{xy}^\text{c}(\vec{k})   & \mathcal{F}_{ \vec{k}}-\Lambda_{\textsc{alt}} & \Omega_{xy}^\text{s}(\vec{k})  & \mathcal{B} \\
		\mathcal{B} & \Omega_{xy}^\text{s}(\vec{k})  & \mathcal{F}_{\vec{k}}+\Lambda_{\textsc{alt}} & -\Omega_{xy}^\text{c}(\vec{k}) \\
		\Omega_{xy}^\text{s}(\vec{k}) & \mathcal{B} &-\Omega_{xy}^\text{c}(\vec{k}) & \mathcal{F}_{\vec{k}}-\Lambda_{\textsc{alt}}
	\end{bmatrix}\!.
\end{equation}
where $\hat{\vec{\Psi}}_{\vec{k}}=[\hat{\psi}_1(\vec{k}),\hat{\psi}_2(\vec{k}),\hat{\psi}^*_1(-\vec{k}),\hat{\psi}^*_2(-\vec{k})]^{\textsc{t}}$ is the 4-component Nambu spinor and $\hat{\eta}=\text{diag}(1,1,-1,-1)$ is a pseudo-Euclidean metric. The set \eqref{eq:Psi-app} has solution $\hat{\vec{\Psi}}=\hat{\vec{\Psi}}_0e^{-i\omega t}$, which turns \eqref{eq:Psi-app} to the eigenvalue problem (EVP)
\begin{equation}\label{eq:EVP}
    \frac{\omega}{\omega_0}\hat{\vec{\Psi}}_0=\hat{\eta}\hat{\mathbb{H}}_{\vec{k}}\hat{\vec{\Psi}}_0.
\end{equation}
This gives the dispersion relation \eqref{eq:omega-main}.

Solving the EVP \eqref{eq:EVP} we compute also the eigenvectors $\hat{\vec{\Psi}}_0(\vec{k})=[\hat{\psi}_{01}(\vec{k}),\hat{\psi}_{02}(\vec{k}),\hat{\psi}^*_{01}(-\vec{k}),\hat{\psi}^*_{02}(-\vec{k})]^{\textsc{t}}$ which enables us to reconstruct dynamics of a magnetic moment sitting in the site $\vec{R}_{\vec{n}}$ when a spin wave with the wave-vector $\vec{k}$ is excited:

\begin{equation}\label{eq:dm-sw}
\begin{split}
    \vec{m}_\nu \approx \vec{m}_\nu^0 + &\sqrt{\frac{2}{N}}  \,\mathrm{Re}\biggl[e^{i(\vec{k}\cdot\vec{R}_{\vec{n}}-\omega_{\vec{k}}t+\varphi_0)} \\
    & \times \left(\vec{e}^+_\nu \hat{\psi}_{0\nu}(\vec{k}) + \vec{e}^-_\nu \hat{\psi}^*_{0\nu}(-\vec{k})\right)\biggr].
\end{split}
\end{equation}
Here $\varphi_0$ is an arbitrary phase. An example of the reconstructed dynamics is shown in Fig.~\ref{fig:singleDyn}.

\section{Magnetic moment of a magnon}\label{app:mu}

To derive the component of the magnetic moment $\mu_\pm^{\text{gs}}(\vec{k})$ along the ground-state N{\'e}el $\vec{n}_0$ vector we introduce an auxiliary small magnetic field along $\vec{n}_0$, i.e., in $x$-direction. Thus the total applied magnetic field is $\vec{B} = B_z\vec{e}_z + B_{x}\vec{e}_x$. Our goal here is to compute the magnon spectrum $\omega_\pm(\vec{k})$ for $B_x\ne0$ and then we define $\mu_\pm^{\text{gs}}$ as
\begin{equation}\label{eq:mu-bx}
    \mu_\pm^{\text{gs}}(\vec{k})=-\hbar\lim\limits_{B_x\to0}\frac{\partial\omega_\pm(\vec{k})}{\partial B_x}.
\end{equation}
In this case, the field component $B_z$ can be finite.

The Zeeman energy contribution is now given by:
\begin{equation}\label{eq:Hz_app}
    \begin{split}
        \mathcal{H}_{\textsc{z}}=&-B_z\mu_s\left[\sum\limits_{\vec{R}_{n}}m_{1z}(\vec{R}_{n})+\sum\limits_{\vec{R}'_{n}}m_{2z}(\vec{R}'_{n})\right] \\
        &-B_{x}\mu_s\left[\sum\limits_{\vec{R}_{n}}m_{1x}(\vec{R}_{n})+\sum\limits_{\vec{R}'_{n}}m_{2x}(\vec{R}'_{n})\right],
    \end{split}
\end{equation}
and the rest of the interactions are the same as defined above. The combination of the longitudinal field $B_z$ and the transverse field $B_x$ induces an asymmetric canting of the sublattice magnetizations ($\theta_1 \neq \theta_2$). Here we assume that $\vec{m}^0_{\nu} = (-1)^{\nu-1}\vec{e}_x \sin\theta_{\nu} + \vec{e}_z \cos\theta_{\nu}$ for $\nu=1,2$. The energy per magnetic unit cell is 
\begin{equation}\label{eq:E_puc_short}
\begin{split}
    E_{\textsc{muc}} = 4J_\textsc{afm} \Big[ \cos(\theta_1 + \theta_2) + \frac{\kappa}{4}(\cos^2\theta_1 + \cos^2\theta_2) \\
    - b_z(\cos\theta_1 + \cos\theta_2) - b_x(\sin\theta_1 - \sin\theta_2) \Big]+\text{const},
\end{split}
\end{equation}
where $b_\alpha = B_\alpha\mu_s / (4J_\textsc{AFM})$ ($\alpha = x,z$) and the constant terms are independent on $\theta_\nu$. Equilibrium values of the canting angles are determined by equations
\begin{equation}
    \begin{split}
    \sin(\theta_1+\theta_2)+\frac{\kappa}{4}\sin2\theta_1-b_z\sin\theta_1+b_x\cos\theta_1=0,\\
    \sin(\theta_1+\theta_2)+\frac{\kappa}{4}\sin2\theta_2-b_z\sin\theta_2-b_x\cos\theta_2=0.
    \end{split}
\end{equation}
In the leading order with respect to $b_x\ll b_z,\kappa$, we obtain $\theta_1\approx\theta_0+\vartheta$ and $\theta_2\approx\theta_0-\vartheta$, where $\vartheta = b_x / [b_z - \frac{\kappa}{2}\frac{\cos2\theta_0}{\cos\theta_0}]$ and $\theta_0$ is the canting angle for $b_x=0$, i.e., $\cos\theta_0 = b_z / (2 + \kappa/2)$. Now we can compute the derivatives
\begin{equation}\label{eq:final_derivatives_short}
    \lim_{B_x \to 0} \frac{\partial\theta_1}{\partial B_x} = -\lim_{B_x \to 0} \frac{\partial\theta_2}{\partial B_x} =\frac{\mu_s}{8J_\textsc{afm}} \frac{\cos\theta_0}{\mathcal{B}+\cos^2\theta_0}.
\end{equation}



Next, using the Holstein-Primakoff representation~\eqref{eq:Tyabl} for the asymmetrically canted ground state, we derive the harmonic part of the Hamiltonian in $k$-space. The different energy contributions are as follows


\begin{widetext}
\begin{equation}\label{eq:Hafm-k-bx-app}	\mathcal{H}_{\textsc{afm}}^{(2)}=4J_{\textsc{afm}}\!\!\sum\limits_{\vec{k}\in\text{1.BZ}}\!\!\left\{-\frac{\cos (\theta_1+\theta_2)}{2}\left[|\hat{\psi}_1(\vec{k})|^2+|\hat{\psi}_2(\vec{k})|^2\right]+\left[\tilde{\Omega}^{s}_{xy}(\vec{k})\hat{\psi}_1(\vec{k})\hat{\psi}_2(-\vec{k})-\tilde{\Omega}^{c}_{xy}(\vec{k})\hat{\psi}_1(\vec{k})\hat{\psi}^*_2(\vec{k})\right]+\text{c.c.}\right\},
\end{equation}
\begin{equation}\label{eq:Han-k-bx-app}
	\mathcal{H}_{\textsc{an}}^{(2)}=\frac{J_\textsc{afm}}{2}\sum\limits_{\nu=1,2}\sum\limits_{\vec{k}\in\text{1.BZ}}\kappa\left[(1-3\cos^2\theta_\nu)|\hat{\psi}_\nu(\vec{k})|^2+\sin^2\theta_\nu \hat{\psi}_\nu(\vec{k})\hat{\psi}_\nu(-\vec{k})+\text{c.c.}\right],
\end{equation}
\begin{equation}\label{eq:Hz-k-bx-app}
	\mathcal{H}_{\textsc{z}}^{(2)}=4J_\textsc{afm}\sum\limits_{\vec{k}\in\text{1.BZ}}\left[ |\hat{\psi}_1(\vec{k})|^2 \left( b_z\cos\theta_1 + b_x\sin\theta_1 \right) + |\hat{\psi}_2(\vec{k})|^2 \left( b_z\cos\theta_2 - b_x \sin\theta_2 \right) \right],
\end{equation}
where we introduce the notations $\tilde{\Omega}^{s}_{xy}(\vec{k})=\Omega_{xy}(\vec{k})\sin^2\frac{\theta_1+\theta_2}{2}$ and $\tilde{\Omega}^{c}_{xy}(\vec{k})=\Omega_{xy}(\vec{k})\cos^2\frac{\theta_1+\theta_2}{2}$. The structure of $\mathcal{H}_{\textsc{alt}}^{(2)}$ and $\mathcal{H}_{\textsc{fm}}^{(2)}$ remain the same as in \eqref{eq:Halt-k-app}, \eqref{eq:Hfm-k-app}.
By following the same procedure as in Appendix~\ref{app:model}, we arrive at the EVP~\eqref{eq:EVP} with the harmonic spin-wave Hamiltonian  
\begin{equation}\label{eq:matrix-H-app}
	\hat{\mathbb{H}}_{\vec{k}}=\omega_0\begin{bmatrix}
		\mathcal{F}_{1 \vec{k}}+\Omega_{1b}(\vec{k}) & -\tilde{\Omega}_{xy}^\text{c}(\vec{k}) & \mathcal{B}_{1}  & \tilde{\Omega}_{xy}^\text{s}(\vec{k})  \\
		-\tilde{\Omega}_{xy}^\text{c}(\vec{k})   & \mathcal{F}_{2 \vec{k}}-\Omega_{2b}(\vec{k}) & \tilde{\Omega}_{xy}^\text{s}(\vec{k})  & \mathcal{B}_{2} \\
		\mathcal{B}_1 & \tilde{\Omega}_{xy}^\text{s}(\vec{k})  & \mathcal{F}_{1 \vec{k}}+\Omega_{1b}(\vec{k}) & -\tilde{\Omega}_{xy}^\text{c}(\vec{k}) \\
		\tilde{\Omega}_{xy}^\text{s}(\vec{k}) & \mathcal{B}_2 &-\tilde{\Omega}_{xy}^\text{c}(\vec{k}) & \mathcal{F}_{2 \vec{k}}-\Omega_{2b}(\vec{k})
	\end{bmatrix},
\end{equation}
\end{widetext}
where $\mathcal{F}_{\nu \vec{k}}=\frac{\kappa}{4} \left( 1-3 \cos^2{\theta_\nu}\right)+\eta\Omega_\textsc{fm}(\vec{k})-\cos{(\theta_1+\theta_2)}+b_z \cos{\theta_\nu}$,  $\Omega_{\nu b}= \varepsilon \Omega_\textsc{alt}(\vec{k})+b_x \sin{\theta_\nu}$, and $\mathcal{B}_{\nu}=\frac{\kappa}{4} \sin^2 \theta_{\nu}$. Now, we compute the eigenvalues $\omega_{\pm}(\vec{k})$ from the EVP~\eqref{eq:EVP} and determine the component $\mu_\pm^{\text{gs}}$ of the magnon magnetic moment using formula~\eqref{eq:mu-bx}. Note that the exact analytical solution for $\theta_1$ and $\theta_2$ is not needed. Since we are interested in the limit $B_x\to0$, it is enough to use Eq.~\eqref{eq:final_derivatives_short} and the fact that $\theta_\nu\to\theta_0$ in this limit. In this way one obtains expression~\eqref{eq:mu-rutile}.

Strictly speaking, in order to prevent the spin flop in the direction of $B_x$, one needs to introduce an auxiliary hard-axial anisotropy $\mathcal{H}'_{\textsc{an}} = K'\left[\sum_{\vec{R}_{\vec{n}}}m_{1y}^2(\vec{R}_{\vec{n}})+\sum_{\vec{R'}_{\vec{n}}}m_{2y}^2(\vec{R'}_{\vec{n}})\right]$ with $K'>0$ and compute the magnon magnetic moment in the limit 
\begin{equation}\label{eq:mu-bx-K}
    \mu_\pm^{\text{gs}}(\vec{k})=-\hbar\lim\limits_{K'\to0}\lim\limits_{B_x\to0}\frac{\partial\omega_\pm(\vec{k})}{\partial B_x}.
\end{equation}
Performing this derivation, we obtain for $\mu_\pm^{\text{gs}}$ the same expression~\eqref{eq:mu-rutile}. So, for the sake of simplicity, we do not introduce the auxiliary anisotropy to the derivation presented above. The analogous derivation of the magnon magnetic moment with the auxiliary anisotropy present can be found in our previous work~\cite{Kravchuk25a}.



\section{Spin-lattice simulations}\label{app:simuls}

The dynamics of magnetic moments are described by the Landau--Lifshits equations
\begin{equation}\label{eq:sim_LL}
        \left(1+\alpha^2_\textsc{g}\right)\frac{\mathrm{d}\vec{m}_{\iota}}{\mathrm{d}t}=\frac{\gamma}{\mu_s}\left(1+\alpha_\textsc{g}\vec{m}_{\iota}\times\right)\vec{m}_{\iota}\times\frac{\partial\mathcal{H}}{\partial\vec{m}_{\iota}},
\end{equation} 
where $\alpha_\textsc{g}$ is the Gilbert damping parameter, $\iota$ enumerates magnetic moments of the lattice, $\mathcal{H}$ is the Hamiltonian used in simulations and defined in Appendix~\ref{app:model}. The dynamical problem is considered as a set of $3N_1 N_2$ ordinary differential equations~\eqref{eq:sim_LL} with respect to $3N_1 N_2$ unknown functions $m_\iota^\textsc{x}(t),\ m_\iota^\textsc{y}(t),\ m_\iota^\textsc{z}(t)$. Parameters $N_1$ and $N_2$ define the size of the system. For the given geometry and initial conditions, the set of time evolution equations~\eqref{eq:sim_LL} is integrated numerically using the Runge--Kutta method in Python.

To simulate the spin waves, we considered a system with a size of $N_1 \times N_2 = 500 \times 500$ magnetic moments. The simulations are carried out in two steps. In the first step, we simulate the dynamics of the system in the low damping regime with $\alpha_\textsc{g}=0.01$ and external magnetic field $\vec{B}_\iota = B_z\vec{e}_z + \tilde{B}_z\vec{e}_z\text{sinc}\left(2\pi\vec{k}\cdot\vec{R}_\iota\right)\text{sinc}\left[2\pi\left(t-t_0\right)\right]$, where $t_0$ is the center of the temporal part of the field profile and $B_z\gg\tilde{B}_z$ are the amplitudes of the applied field.

In the second step we perform a space-time Fourier transformation for the complex-valued parameter $m_\iota^\textsc{x}(t)+i\, m_\iota^\textsc{y}(t)$. The resulting eigenfrequencies are plotted in Fig.~\ref{fig:spec}.

\section{Spin conductivity tensor}\label{app:sigma}

Going from summation over the Brillouin zone to integration $\frac{1}{L_xL_y}\sum_{\vec{k}}(\dots)\to\frac{1}{(2\pi)^2}\int(\dots)d\vec{k}$ and introducing the dimensionless wave vector $\vec{q}=a_0\vec{k}$, eigenfrequency $\tilde{\omega}_{\vec{q},\nu}=\omega_\nu(\vec{q}/a_0)/\omega_0$, group velocity $\tilde{\vec{v}}_{\vec{q},\nu}=\partial_{\vec{q}}\tilde{\omega}_{\vec{q},\nu}$, and magnetic moment $\tilde{\mu}^{\text{gs}}_{\vec{q},\nu}=\mu^{\text{gs}}_{\vec{q}/a_0,\nu}/(|g|\mu_{\textsc{b}})$, we write the out-of-diagonal component in form $\sigma_{xy}=\sigma_0\varsigma(k_{\textsc{b}}T/(\hbar\omega_0))$, where $\sigma_0=\tau_{\text{rlx}}\omega_0|g|\mu_{\textsc{b}}/\hbar$ and
\begin{equation}\label{eq:sigma-dmnls}
    \varsigma=-\frac{1}{\pi^2}\!\sum\limits_{\nu=\pm}\iint\limits_0^\pi\!\!\frac{\tilde{\mu}^{\text{gs}}_{\vec{q},\nu}(\tilde{v}_{\vec{q},\nu})_x(\tilde{v}_{\vec{q},\nu})_y}{\tilde{\omega}_{\vec{q},\nu}}\left[\frac{\tilde{\beta}\tilde{\omega}_{\vec{q},\nu}/2}{\sinh\frac{\tilde{\beta}\tilde{\omega}_{\vec{q},\nu}}{2}}\right]^2\!\!\!\dd q_x\dd q_y.
\end{equation}
Here $\tilde{\beta}=\hbar\omega_0/(k_{\textsc{b}}T)$ and we utilized the symmetries of $\tilde{v}_{\vec{q},\nu}$, $\tilde{\mu}^{\text{gs}}_{\vec{q},\nu}$ and $\tilde{\omega}_{\vec{q},\nu}$ discussed in the main text. 

In the high-temperature limit $\tilde{\beta}\ll1$, the bracket in the integrand in~\eqref{eq:sigma-dmnls} approaches 1, therefore the spin conductivity in this regime
\begin{equation}\label{eq:sigma-xy-HT}
    \sigma_{xy}^0\approx-\frac{\sigma_0}{\pi^2}\sum\limits_{\nu=\pm}\iint\limits_0^\pi\frac{\tilde{\mu}^{\text{gs}}_{\vec{q},\nu}(\tilde{v}_{\vec{q},\nu})_x(\tilde{v}_{\vec{q},\nu})_y}{\tilde{\omega}_{\vec{q},\nu}}\dd q_x\dd q_y
\end{equation}
is temperature-independent.


%

\end{document}